# Au Nanoparticles in Lipid Bilayers: a Comparison between Atomistic and Coarse Grained Models

*Sebastian Salassi[1], Federica Simonelli[1], Davide Bochicchio[1], Riccardo Ferrando[2] and Giulia Rossi*[1]*

[1] Physics Department, University of Genoa, via Dodecaneso 33, 16146 Genoa, Italy

[2] Chemistry Department, University of Genoa, via Dodecaneso 31, 16146 Genoa, Italy

*rossig@fisica.unige.it

## Abstract

The computational study of the interaction between charged, ligand-protected metal nanoparticles and model lipid membranes has been recently addressed both at atomistic and coarse grained level. Here we compare the performance of three versions of the coarse grained Martini force field at describing the nanoparticle-membrane interaction. The three coarse-grained models differ in terms of treatment of long-range electrostatic interactions and water polarizability. The NP-membrane interaction consists in the transition from a metastable NP-membrane complex, in which the NP is only partially embedded in the membrane, to a configuration in which the NP is anchored to both membrane leaflets. All the three coarse grained models provide a description of the metastable NP-membrane complex that is consistent with that obtained using an atomistic force field. As for the anchoring transition, the polarizable-water Martini correctly describes the molecular mechanisms and the energetics of the transition. The standard version of the Martini model, instead, underestimates the free energy barriers for anchoring and does not completely capture the membrane deformations involved in the transition process.

**Introduction**

The use of monolayer-protected inorganic nanoparticles (NPs) as target-selective drug vectors[1,2], nanothermal[3-6] agents or diagnostic devices[7,8] requires that we achieve control on the NP interaction with different biological environments. The interaction of NPs with cell membranes, in particular, is crucial for the delivery of NPs into cells, and is the subject of intense research efforts aimed at understanding the molecular basis of active, endocytic internalization pathways[9] as well as of passive membrane permeation. Here we focus on the latter mechanism, which has been shown to be relevant for the smallest NPs (diameter < 10 nm) interacting with plasma membranes and model lipid bilayers[10-13].

Membrane passive translocation rates are the result of a complex interplay of thermodynamics and kinetics. From a thermodynamic point of view, the degree of hydrophilicity of the NP determines its propensity to reside in the water phase or in the hydrophobic membrane core[11,14,15]; as it is often the case, the flexible NP ligand shell can make the NP quite adaptable to the surrounding environment[14], leading to the stabilization of long-lived metastable configurations both in the extra or intracellular water environment and in the membrane core[16,17]. The kinetic availability of transition pathways and the free energy barriers between these metastable states eventually determine passive permeation rates[18-20].

Recently, a series of experimental papers[11-13,21,22] have focused on the study of a family of charged, monolayer-protected Au NPs and on their interactions with plasma membranes and model lipid bilayers. These NPs are functionalized by a mixture of hydrophilic, negatively charged ligands (mercapto undecane sulphonate, $-S-(CH_2)_{11}-SO_3^-$ (MUS)), or mercapto undecane carboxylate, $-S-(CH_2)_{11}-CO_2^-$ (MUC)) and neutral, hydrophobic ligands (octanethiol, $-S-(CH_2)_7-CH_3$ (OT)) and they are small enough to allow passive membrane translocation (diameter 2 or 4 nm). Neutron reflectivity data[22] indicate that they can interact in a non-destructive way with the surface of floating zwitterionic bilayers, and confocal microscopy observations[11] show that they can be co-localized with the bilayers of multilamellar vesicles without causing any leakage. The same NPs were shown to passively penetrate the plasma membrane of HeLA cells[12], where their propensity to follow the passive permeation pathway might depend on the spatial arrangement of ligands on the NP surface.

The computational approach to the study of NP-membrane interactions can complement the experimental investigation as it has the advantage of offering an atomistic or at least molecular interpretation of the permeation mechanism. Unfortunately, membrane permeation rates for NPs can easily span time scales of seconds that are not currently within reach for unbiased atomistic Molecular Dynamics simulations. One possible strategy to overcome the sampling barrier is to rely on coarse-grained models that couple a reduction of the system degrees of freedom to an intrinsically faster dynamics. Indeed, atomistic[16,20,23–25] and coarse-grained[15,19,18] molecular dynamics simulations have nicely complemented each other in the recent literature[26], converging on the study of the same anionic, MUS/MUC- and OT-passivated Au NPs and eventually proposing a three-stage mechanism of NP-membrane interaction. According to the atomistic simulations of Heikkilä[16,23] *et al.*, anionic Au NPs could stably adhere to the surface of zwitterionic lipid bilayers as a result of favorable electrostatic interactions between their charged ligands and the polar lipid headgroups; then, according to both atomistic[20] and coarse-grained[18] simulations, NPs would partially penetrate the membrane by establishing a hydrophobic contact (HC) between the hydrophobic moieties of their ligands and the lipid tails of the entrance leaflet; eventually, NPs would find their way towards a so-called snorkeling or anchored configuration, in which the NP charged ligands interact with the lipid headgroups of both membrane leaflets. The stability of the anchored configuration had been predicted by implicit solvent models[27], too. The transition from the hydrophobic contact configuration to the anchored state has a slow kinetics, and it has been shown to occur spontaneously on flat membranes only by the coarse-grained simulations of Simonelli[18] *et al.*, while the atomistic simulations of Van Lehn *et al*. could simulate the process in presence of a highly-curved membrane[25]. Figure 1 sketches the three main stages of the NP-membrane interaction.

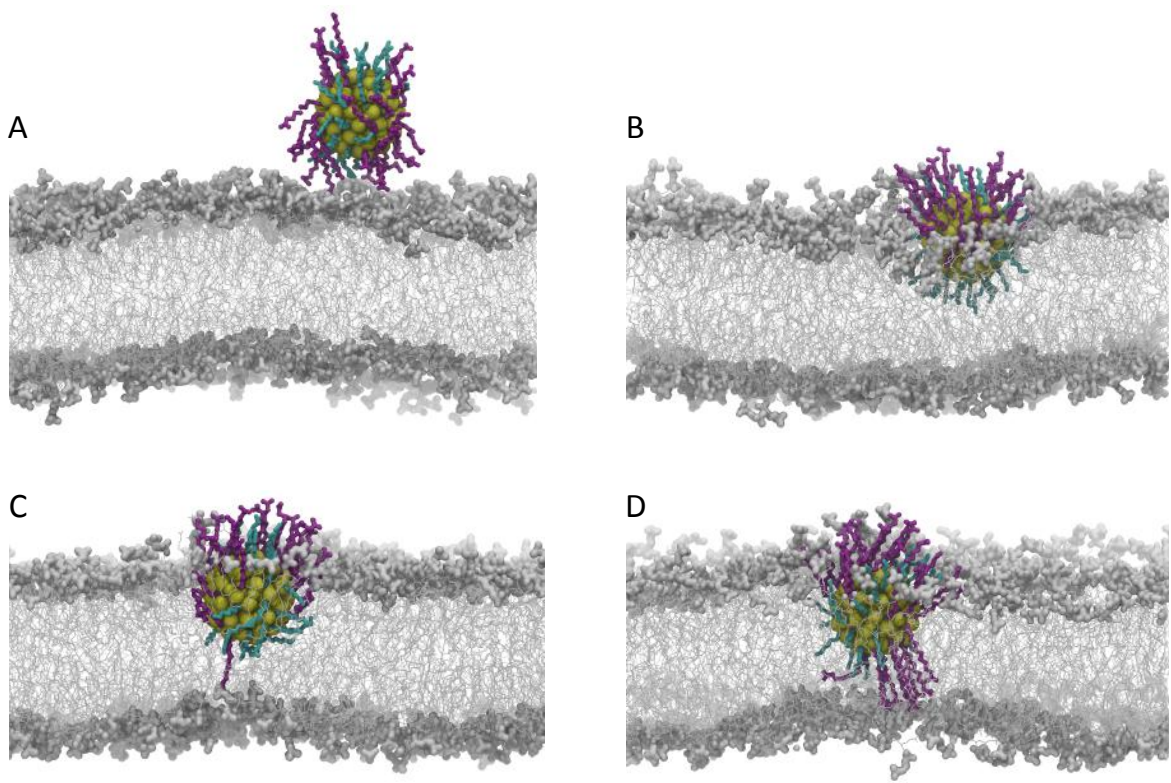

**Figure 1 - The proposed mechanism for the NP-membrane interaction.** Palmitoyloleoyl phosphatidylcholine (POPC) lipids in grey, with the headgroups in spacefill representation. Au-S NP core in yellow, hydrophobic OT ligands in cyan, anionic MUC ligands in purple. A: the NP is adsorbed at the membrane surface; B: the NP is in the HC configuration, only partially embedded in the bilayer, with the many hydrophobic NP ligands in contact with the lipid tails, and the anionic ligand terminals in contact with the lipid headgroups of the entrance leaflet. C and D: the NP progressively anchors to the distal leaflet by dropping one ligand (C) after the other until a snorkeling configuration is reached (D). (These atomistic snapshots are purely representative of the proposed mechanism and were not derived from any of the simulations analyzed in this work).

While these results are a nice example of the possible convergence and complementarity of molecular simulations performed at different resolutions, the computational approach still faces hard challenges. Enhanced sampling techniques are often used to accelerate the sampling of rare translocation events, but they still rely on a subtle assessment of the relevant reaction coordinates. The choice of the appropriate reaction coordinate can be challenging for the permeation of rather small solutes already[28], and it becomes even harder when looking at the permeation of ligand-protected NPs with very large conformational flexibility. This latter issue is common to atomistic and coarse-grained approaches. Eventually, an appropriate and quantitative description of the charged NP-membrane interaction depends crucially on the ability of the force field to reproduce the correct solvation free energies of the charged moieties that are transferred between water and the membrane core. At atomistic level, there is general agreement on the height of the free energy barriers for the membrane translocation of monovalent ions[29,30], while coarse-grained force fields such as the Martini[31] force field, previously used to study anionic NP-membrane interactions[15,18,19,32], can severely underestimate them[33].

In this paper we test the performance of three versions of the popular coarse-grained Martini force field at reproducing the atomistic free energy profile associated to the first step of the anchoring transition (from Figure 1B to Figure 1C). This interaction step consists in the translocation from the entrance to the distal leaflet of one single charged ligand terminal. The three models we consider are the standard version[31] of the Martini (SM) force field, in which electrostatic interactions are treated as short-range interactions, with a dielectric constant $\varepsilon = 15$; the standard Martini force field modified by the inclusion of long-range electrostatics, implemented via Particle-Mesh-Ewald summation (MPME); and the polarizable Martini force field (MPW) that treats electrostatic interactions with a dielectric constant $\varepsilon = 2.5$ and includes long range electrostatics and water polarizability[33].

We find that, contrary to the SM force field, the MPW force field provides estimates of the translocation barrier that are close to those predicted by our atomistic calculations. Moreover, the MPW force field is the most accurate at reproducing the molecular mechanisms involved during the anchoring transition as predicted by atomistic simulations.

**Methods**

*Atomistic model*. We set up an atomistic united-atom (UA) model of an anionic, MUC- and OT-functionalized Au NP compatible with the OPLS force field[34] and with the Berger parameters for lipids[35]. The core of the NP, as derived by Lopez-Acevedo[36] *et al.*, is made of 144 Au atoms with icosahedral symmetry and 60 S atoms bound to the gold core. Both Au and S atoms are connected through an elastic network.

A total of 60 ligands are bound to the NP core via Au-S bonds. The ligands are 30 hydrophobic OT and 30 negatively charged mercapto-undecane carboxylate (MUC). The chemical structure of the ligands is shown in Fig. S1. The parameterization of the Au core and of the two ligands, including bonded and non-bonded parameters, is summarized in the Supporting Information.

*Coarse grained model.* The description of the core of the NP is the same as in the atomistic model, except for charges – all Au and S atoms are neutral in the CG model. Au and S van der Waals interactions are short range and purely repulsive. OT is made of 2 connected Martini beads of type C1 while MUC is built with 3 beads of type C1 and 1 terminal Qda bead. Bond and angle parameters along the ligands are the same as for Martini alkanes. Further details are reported in the Supporting Information. Here we will consider only patched NPs, in which ligands with azimuthal angle within a certain range are hydrophobic while all other ligands are charged, giving rise to a "patched" arrangement.

*Simulation set up for the unbiased MD runs.* In our atomistic simulations we set up an initial configuration in which the NP is in the hydrophobic contact state (Fig.1B and Section 3 of the SI). The membrane is composed of 480 zwitterionic POPC lipids (13.2×13.2 nm), solvated in a box of ~32000 water molecules and 30 positive ions to compensate the charge of the AuNP. Salt was added to the solvent at physiological concentration (150 mM).

The initial configuration for the CG unbiased runs is built as in the atomistic case, with a membrane composed of 512 POPC lipids (13.6×13.6 nm). The NP-membrane complex is solvated with ~15320 Martini water beads. We performed different equilibration runs for the SM and the MPME models. For the MPW model, the equilibration run was preceded by the conversion of the standard water beads to the polarizable water beads. In all cases, 30 Na+ ions were included to neutralize the system.

Simulation parameters, including equilibration times, are summarized in Table 1.

*Metadynamics simulations.* The process under study with metadynamics[37,38] simulations is the reversible anchoring-disanchoring transition of the biased ligand terminal. The anchoring, or forward process consists in the translocation of one charged ligand across the membrane (from the hydrophobic contact state to the anchored state); the disanchoring, or backward process consists in the transition of the same ligand back to the starting configuration. The collective variable, $\zeta$, used in our CG and UA metadynamics simulations is the distance along the z-axis, perpendicular to the membrane plane, between the center of mass (COM) of the membrane and the COM of the biased charged terminal group. The starting configurations of our metadynamics runs, in which the NP is in the HC state, were extracted from the unbiased MD simulations. We have chosen the ligand to be biased as the one whose covalent link to the NP core had the lowest $z$ coordinate in the initial configuration (see also Fig.S3). The other run parameters are listed in Table 1.

A first set of metadynamics runs, both atomistic and CG, were run until a complete forward + backward cycle had been performed (Table 1, third to last line). The data collected during the whole runs were used to analyze the structural features of the membrane during the forward and backward transitions. In order to quantify the energy barriers associated to the forward process, we followed a standard procedure, as described by Laio and Gervasio[38], and we analyzed the complete metadynamics trajectories until the time at which the forward transition had been completed (Table 1, second to last line). More details about the way in which we identified the transition are reported in Section 6 of the SI (Fig.S5). We ran a second set of independent metadynamics runs, at CG level only, to quantify the barriers of the backward process (Table 1, last line). These runs were initialized with different frames extracted from the complete metadynamics simulations with the NP in the anchored state.

The potential of mean force (PMF) for the anchoring and the disanchoring processes was obtained averaging the PMFs of several independent metadynamics runs (the total simulation time and number of runs are shown in Table 1). The error reported in the PMF plots is the standard error.

*Contact analysis*. The number of contacts between the various components of the NP, water and lipid choline groups were obtained using the *mindist* Gromacs tool with a threshold contact distance of 0.6 nm.

All MD simulations were performed with Gromacs[39] v5 patched with Plumed[40] 2.

Table 1 Simulation parameters. In metadynamics runs, $\tau$ is the time interval between two consecutive depositions of the Gaussian bias, $h$ and $\delta w$ are the Gaussian height and width, respectively.

| MD settings | Atomistic | Coarse grained |
|---|---|---|
| Time step | 2 fs | 20 fs |
| P coupling (equilibration) | Berendsen (1 bar) | idem |
| P coupling (production) | Parrinello—Rahman (1 bar) | idem |
| T coupling | Velocity-rescale (310 K) | idem |
| PME grid spacing | 0.20 nm | 0.12 nm |
| Unbiased MD runs | | |

| Equilibration time | 30 ns | 1 μ |
|---|---|---|
| Total production run time (no. of runs) | 500 ns | SM: 10 μs (1) <br> MPW: 10 μs (1) |
| **Metadynamics runs** | | |
| $\tau$ | 0.1 ns | 1 ns |
| $h$ | 1 kJ/mol | 2.48 kJ/mol |
| $\delta w$ | 0.06 nm | *idem* |
| Forward + backward (no. of runs) | 1.9 μs (2) | SM: 3.1 μs (8) <br> MPME: 5.2 μs (10) <br> MPW: 16.6 μs (10) |
| Forward only (no. of runs) | | SM: 1.1 μs (8) <br> MPME: 2.1 μs (10) <br> MPW: 10.3 μs (10) |
| Backward only (no. of runs) | | SM: 1.3 μs (8) <br> MPME: - (0) <br> MPW: 4.5 μs (7) |

**Results**

*A. Structural characterization of the hydrophobic contact stage*

Our results based on the use of the SM model[18] had shown that the hydrophobic contact configuration (Figure 1B, indicated by HC) can be stable during unbiased MD. We showed also that the typical life time of the HC configuration can depend on the spatial arrangement of ligands on the surface of the NP. Here we will consider only patched NPs, whose surface is characterized by a central stripe of hydrophobic OT ligands, flanked by two stripes of MUC (or MUS) ligands (Figure S2). This configuration allowed for the longest stabilization of the HC state with the SM model, with recorded life times of 5 ns, 3.7 μs, 3.9 μs and 9.3 μs before the anchoring transition spontaneously took place. In one case, no spontaneous anchoring occurred after 10.8 μs. When switching to atomistic, MPME and MPW models, we could never observe spontaneous transitions during the simulated time (see the unbiased MD runs in Table 1). Other arrangements of the ligands on the surface of the NP are possible and can influence the NP-membrane interaction mechanisms. In Simonelli[18] *et al.* for example, we showed that a random arrangement of the ligands on the NP surface can reduce the life time of the HC configuration, while leaving unaltered the overall NP-membrane interaction mechanism. Here we limit our investigation at the analysis of patched NPs, with the aim to compare the performances of different models on the same benchmark system.

In the HC state, the NP partially penetrates the membrane. In Table 2 we report the $z$ component of the distance between the center of mass of the NP and the center of mass of the POPC membrane, $d^z_{COM-COM}$ as calculated with the different models[1]. In this respect, the MPW model is in excellent agreement with the atomistic one, with $d^z_{COM-COM} \sim 1.7$ nm, while the other CG models favor a configuration in which the NP is more deeply inserted in the membrane. This is reflected also by the average number of contacts between the hydrophobic beads of the NP and the hydrophobic beads of the lipid tails, which sets at $301 \pm 1$ for the SM and at $286 \pm 1$ for the MPW model. The average number of contacts between the charged ligand terminals and water beads is $135 \pm 1$ for the SM and $160 \pm 1$ for the MPW.

**Table 2 Average distance between the COM of the NP and the COM of the POPC membrane in the HC state**

|  | Atomistic model | SM | MPME | MPW |
|---|---|---|---|---|
| $d^z_{COM-COM}$ [nm] | $1.708 \pm 0.008$ | $1.463 \pm 0.002$ | $1.554 \pm 0.005$ | $1.717 \pm 0.002$ |

The presence of the NP alters the membrane structure in several ways. In the entrance leaflet, the density of the polar lipid heads is increased around the NP, both in the atomistic and in the CG simulations (Figure S6, top left). The presence of the NP in the HC state affects the structure of the distal leaflet as well. The lipid heads of the distal leaflets are denser below the NP (Figure 2), with a 12% increase in atomistic and SM runs, and a larger 20% increase in MPW runs. In CG simulations we also observe an increase of the density of lipid tails around the NP, both in the entrance and in the distal leaflet (Figure S6, bottom panels) an effect that is almost absent in the atomistic runs. 2D plots of the lipid heads and tails densities are shown in S7 and S8, for the MPW and atomistic models, respectively.

---

[1] Here and in the following, the $z$ components of the distances calculated in the CG simulations will be rescaled by the ratio between the thickness of the POPC membranes in the atomistic and in the CG model.

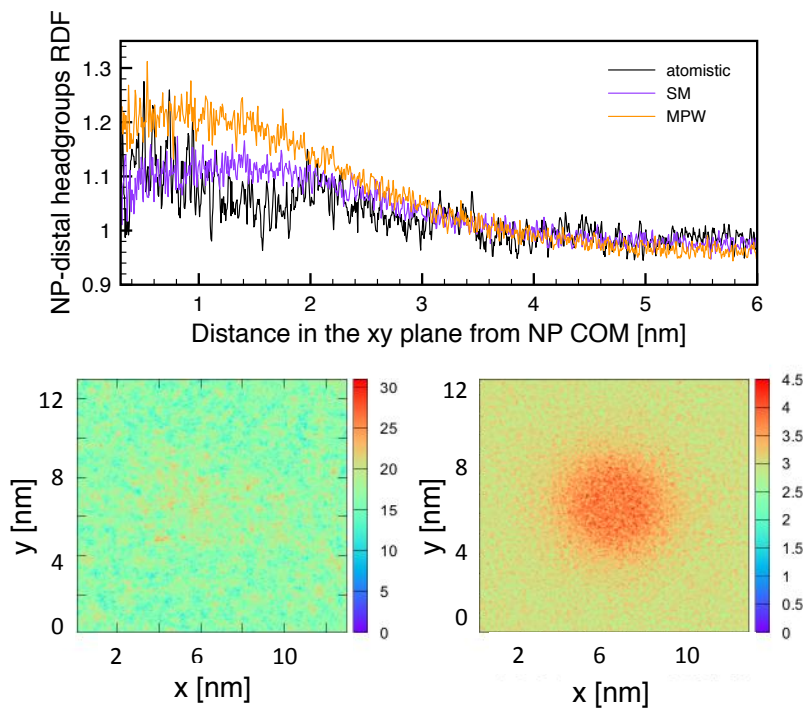

**Figure 2** Top: In-plane radial distribution functions of the phosphate and choline groups in the distal leaflet, from the COM of the NP in the HC state. Bottom: 2D maps of the numeric densities of lipid heads in the distal leaflet as obtained with the atomistic (left) and MPW (right) model.

## B. Structural and dynamical characterization of the HC state→anchored state transition

As no spontaneous transition from the HC configuration to the anchored state was observed during the atomistic, MPME and MPW runs, our comparison of the transition mechanisms observed with the different models is based on the forward + backward metadynamics runs (as listed in Table 1).

In Figure 3 we plot the average number of contacts of the biased charged ligand terminal with the choline groups of the lipid heads of the entrance leaflet (Figure 3). For $\zeta > 2.5$ nm, corresponding to configurations in which the biased ligand terminal is located above the lipid head region, in the water phase, the atomistic and CG data are in reasonable agreement. As the ligand terminal approaches the center of the membrane, though, the four models predict different scenarios. In the atomistic runs, the number of contacts with water is non negligible even at the center of the membrane. In the CG runs, instead, only the MPW model is able to reproduce this feature, still underestimating the number of water molecules in contact with the translocating ligand.

We also monitored the electrostatically favorable contacts between the biased ligand and the choline groups of the entrance leaflet (Figure 3). In the atomistic runs the biased ligand interacts with the choline groups over a broader $z$ range, including when it explores the distal membrane leaflet. All the CG models, and especially the MPW model, appear to overestimate the number of ligand-choline contacts. The MPW model shows a slightly wider $z$ range in which ligand-choline contacts are recorded, reaching down to the center of the membrane.

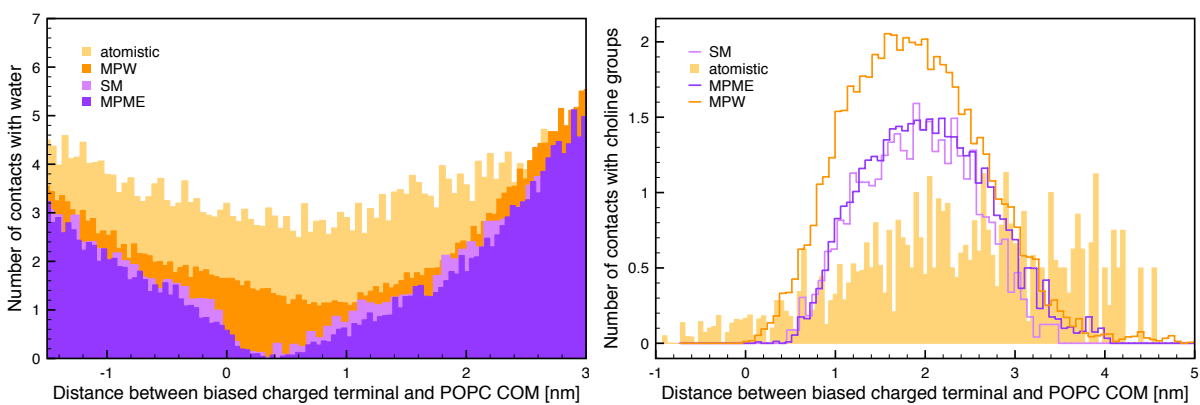

**Figure 3 – Number of contacts between the biased charged ligand terminal and the water groups (left) and choline (right). In order to compare atomistic and CG data, the number of contacts with water molecules in the atomistic run has been divided by 4 to account for dimension of the CG beads.**

The data shown in Figure 3 are the result of a time average over the whole forward + backward process. The biased ligand explores the central region of the membrane, both coming from the HC and from the anchored state. During these excursions in the membrane core, the ligand always preserves some contacts with water molecules and choline groups. In Figure 4 we show the time evolution of $\zeta$ in a few hundreds ns before the forward transition. The time evolution of $\zeta$ in the atomistic runs shows strong correlations with the minimum distance between the choline groups and the COM of the membrane, as well as with the minimum distance between water and the COM of the membrane. This correlation is almost absent in the SM runs, while it is correctly reproduced by the MPW model. The biased ligand drags water and lipid headgroups towards the center of membrane while attempting the transition to the anchored state. This is accompanied by important, though local, membrane deformations. Examples of such deformations are shown in Figure 5.

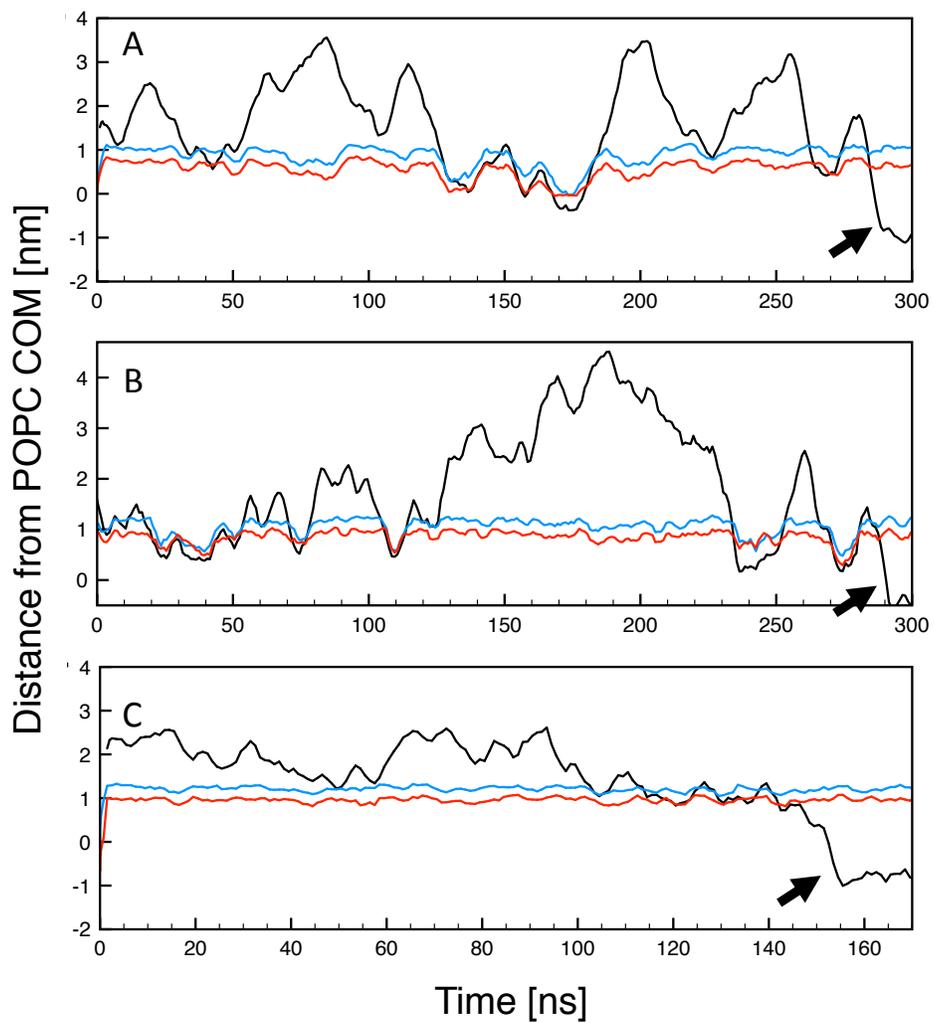

**Figure 4** – Time evolution of the collective variable ζ (black line), of the minimum distance between the choline groups of the entrance leaflet and the COM of the membrane (blue), and of the minimum distance between water and the COM of the membrane (red). A, B and C panels refer to atomistic, MPW and SM metadynamics runs, respectively. The arrows indicate the transition to the anchored configuration.

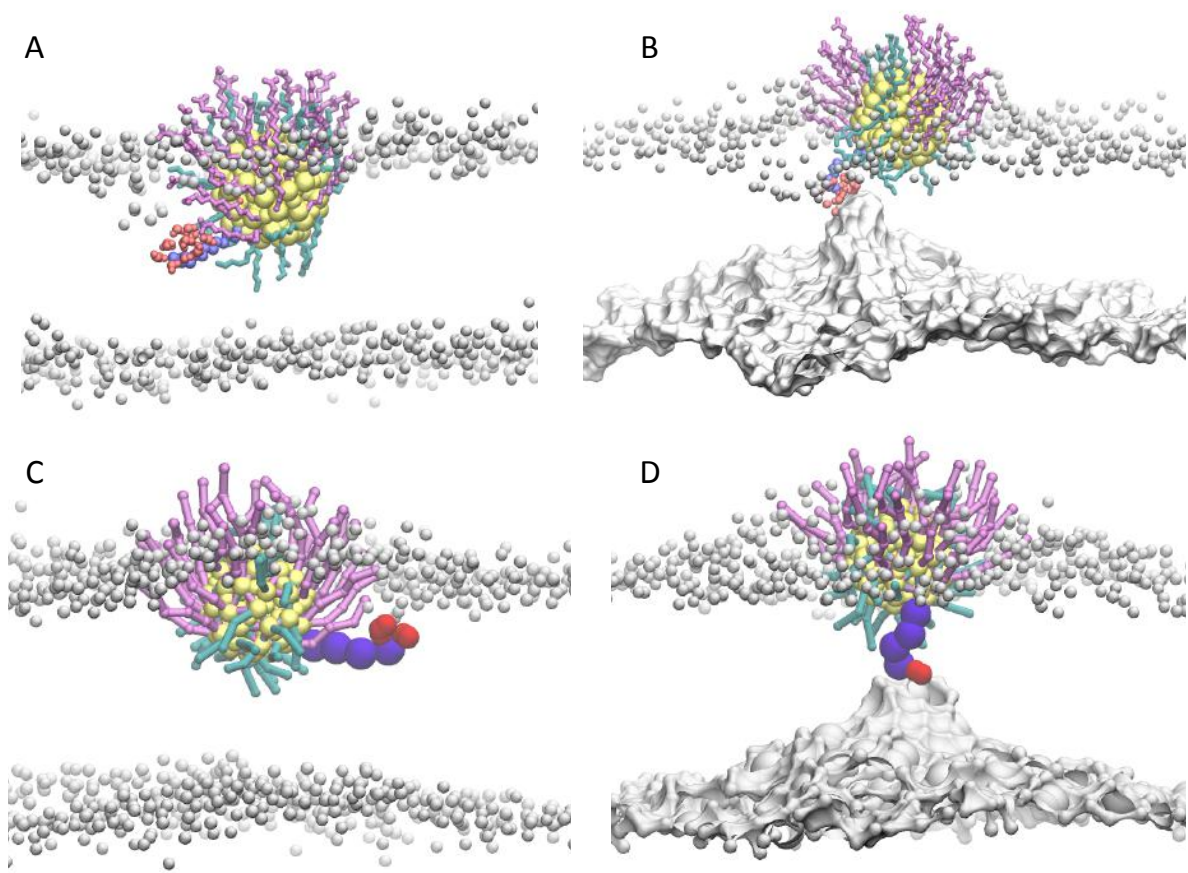

**Figure 5** – Lipid heads in grey, tails not shown. Au and S atoms in yellow. Hydrophobic ligands in cyan, anionic ligands in pink. Water in red. A and B: snapshots from an atomistic metadynamics run. A shows the biased anionic ligand (violet) exploring the central part of the membrane while dragging water molecules (red) towards the membrane core. In B, the biased ligand is anchored to the distal leaflet and deforms it while wandering in the central part of the membrane (the lipid heads of the distal leaflets are represented by a continuous surface to better visualize the membrane deformation). C and D: similar configurations from a CG metadynamics run performed with the MPW model.

During the actual transition to the anchored state (last ~10 ns of the plots in Figure 4) and back to the HC state we monitored whether any water molecule crosses the membrane core together with the biased ligand, leading to the formation of single water files or bulk water pores[41,29]. Our 2 atomistic metadynamics simulations show that different mechanisms can be activated. In one forward transition (Figure 6, top row) the ligand translocates without dragging any water molecule to the distal leaflet. The other forward transition (Figure 6, bottom rows) is accompanied by the formation of a water pore, whose life time is 2-3 ns. Several water molecules pass from the entrance to the distal leaflet in this case. During the backward transitions (Figure S9), we have observed both the formation of a water pore and of a single water file. At CG level, no pore formation has been ever observed. In the metadynamics performed with the SM we have never recorded any water molecules crossing the membrane core during the ligand translocation, neither during the forward nor during the backward transition. In the metadynamics runs performed with the MPME model, 2 out of 20 transitions (Table 1) were characterized by the translocation of a single water bead. With the MPW, instead, in 9 out of 20 transitions (Table 1) the ligand transferred a single water bead through the membrane core.

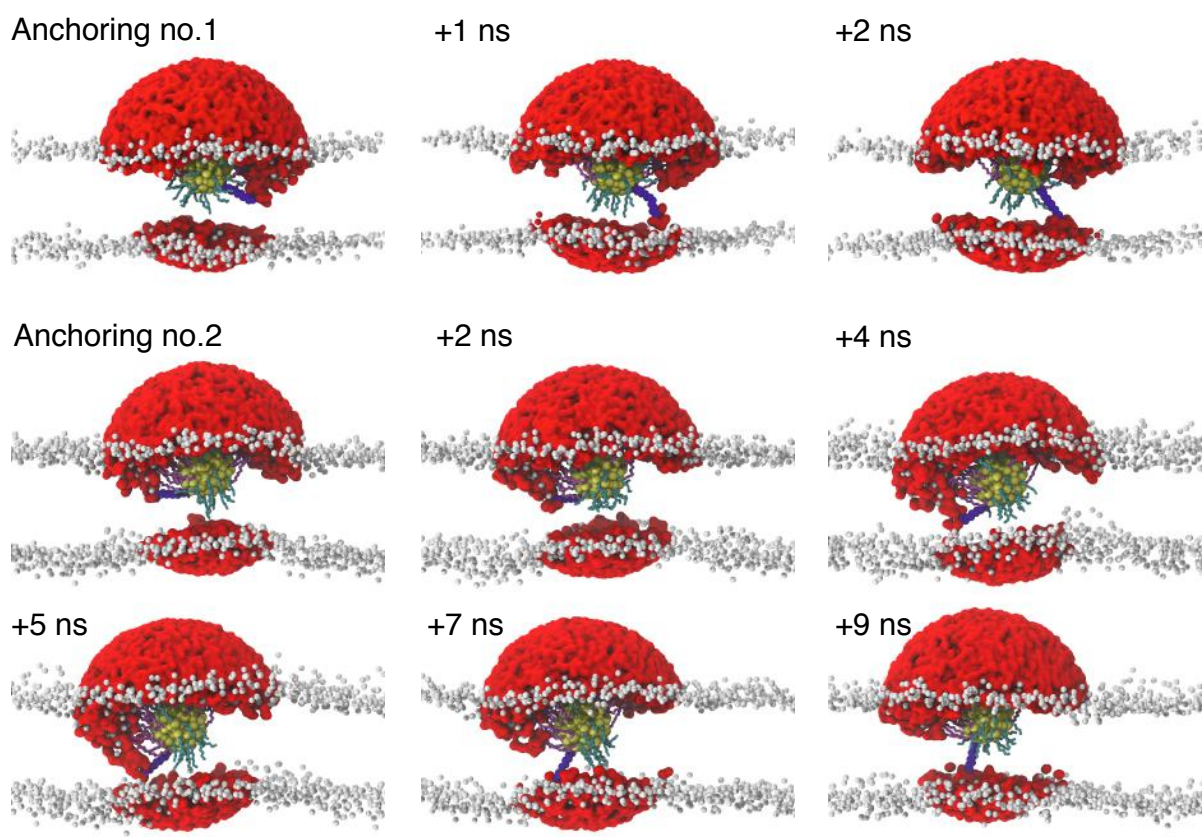

**Figure 6** – Snapshots from the atomistic metadynamics runs showing the molecular mechanisms involved during the biased ligand anchoring. Colors are as in Figure 5. All water molecules within 3.5 nm from the NP center of mass are shown in red. In the first run (top row), the formation of a water defect is followed by the forward transition without passage of water molecules through the membrane core. In the second run (bottom rows), the anchoring is accompanied by the formation of a water pore.

## C. PMFs for anchoring/disanchoring

In this section we focus on the quantification of the free energy barriers that the biased ligand has to overcome to perform the forward, anchoring transition and the backward, disanchoring transition. The metadynamics runs we used to this scope are reported in Table 1. The potential of mean force (PMF) for the forward process was derived from the complete metadynamics runs (forward + backward), truncated as soon as the forward transition had been completed. The PMF for the backward process was derived from independent metadynamics runs initialized in the anchored state and interrupted as soon as the ligand had come back to the entrance leaflet.

The forward barrier, $\Delta E_f$, according to our reference atomistic runs, is extremely large: 135 kJ/mol. Based on the comparison of our two independent runs, the uncertainty of the atomistic data is about $\pm 10$ kJ/mol. The SM severely underestimates the free energy barrier for the forward process, which is $26 \pm 3$ kJ/mol. The addition of long-range electrostatics does not improve much the comparison with the atomistic model, raising the barriers up to $36 \pm 5$ kJ/mol. The MPW model, instead, provides a barrier of $100 \pm 8$ kJ/mol, much more in line with the atomistic result.

We compared the backward barriers, $\Delta E_b$ as predicted by the SM and MPW models, as well. According to the SM model, the backward process is disfavored with respect to the forward one[18], with a barrier of $38 \pm 5$ kJ/mol. The MPW model, instead, predicts a backward barrier of $101 \pm 7$ kJ/mol, thus attributing no thermodynamic advantages to any of the two states.

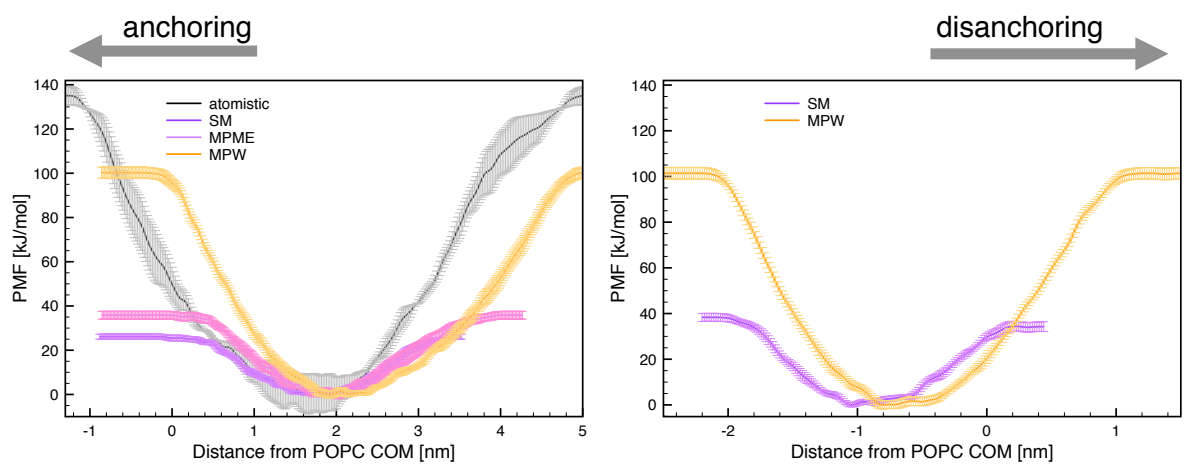

**Figure 7 – PMFs for the forward anchoring process (left) and for the backward disanchoring process (right). Error bars are standard errors from the set of metadynamics runs listed in Table 1.**

Table 3 – Free energy barriers for the anchoring and disanchoring transition of a single charged ligand, as predicted by atomistic and CG simulations. Energies in [kJ/mol].

|  | Atomistic model | SM | MPME | MPW |
|---|---|---|---|---|
| $\Delta E_f$ | 135 ± 10 | 26 ± 3 | 36 ± 5 | 100 ± 8 |
| $\Delta E_b$ | - | 38 ± 5 | - | 101 ± 7 |

**Discussion**

*The stability of the HC configuration according to the different CG models.*

Our previous unbiased simulations of the NP-membrane complex, performed with the SM model, had predicted that the HC configuration could be stabilized for several $\mu$s, before the spontaneous transition to the anchored configuration was observed. When switching to the MPME or MPW models, none of our unbiased simulations led to a spontaneous transition to the anchored state in 10 $\mu$s, suggesting that the introduction of long-range electrostatics, of polarizable water and of a more realistic treatment of the dielectric properties of the membrane core have the overall effect of stabilizing the HC state.

In the HC configuration the NP alters some of the membrane structural properties, inducing variations of the lipid densities both in the entrance and in the distal leaflet. Overall, the CG models show some tendency to overestimating these density fluctuations (Figure 2, Figure S6, S7 and S8), with the SM and MPME models sometimes in better agreement with the atomistic force field than the MPW model (see Figure 2).

*Adding PME to the standard Martini does not significantly change NP-membrane interactions.*

The MPME model has been used in the past to simulate the interaction of other charged molecules with model lipid bilayers. The first stages of the interaction between positively charged dendrimers and model lipid bilayers, for example, closely resembles the anchoring of the charged Au NP ligands; Lee and Larson[42,43] have shown that the addition of long-range electrostatic interactions to the standard Martini model can speed-up the dendrimer interaction with the membrane, leading to the formation of toroidal pores within the simulated time (< 1 $\mu$s). Similar toroidal pore formation in bilayers can be induced by antimicrobial peptides, as observed by Rzepiela[44] *et al*. again with the MPME model. In the latter study, the authors state that the introduction of PME stabilizes the toroidal pore structure, with no major effects on the membrane structure and dynamics in absence of the peptide. In our simulations, the use of PME does not cause any major modification to the membrane structure when the NP is in the HC configuration. The equilibrium distance between the center of mass of the NP and the membrane center is only slightly larger than in the SM simulations, while the radial distribution of lipids around the NP is unchanged. The MPME model predicts a free energy barrier for the anchoring transition that is about 10 kJ/mol larger than for the SM model; this can be rationalized in terms of strengthening of the interactions between the charged ligands and the lipid headgroups. Still, the free energy barrier is about 70 kJ/mol smaller than that calculated by the atomistic metadynamics.

*The polarizable water Martini model predicts a much larger anchoring barrier and a ligand-induced defect mechanism for the NP anchoring.*

While the free energy barrier for the translocation of a single charged ligand is clearly underestimated by the SM and MPME models, the MPW is in much better agreement with the atomistic result (Table 3). Moreover, the MPW model seems to reproduce more accurately the molecular mechanisms involved during the anchoring process. The membrane deformations we observe during the attempts of the biased ligand to cross the membrane core are quite similar to that reported for polar or charged aminoacids[45–48]. This mechanism is also the one that is believed to be the most likely for the translocation of single ions[47,30] where the direct permeation has been attributed to the formation of ion-induced defects involving the dragging of water molecules and lipid headgroups towards the center of the membrane. In our atomistic metadynamics runs, 3 over 4 ligand translocations (forward and backward) happen together with the formation of a water pore or of a single water file spanning the membrane thickness. The MPW model, with 9 out of 20 transitions accompanied by the translocation of one water bead, are in line with the atomistic result, taking into account the steric effects limiting the penetration of the CG water in the lipid tail region.

Our first conclusion is that, when looking at the interaction of charged NPs with lipid membranes, the SM model remains a valid tool to screen for possible interaction mechanisms in a time-effective way. A more accurate description of electrostatics at CG level is nevertheless essential to quantify translocation barriers and describe, as correctly as possible with the CG resolution, the membrane deformations during the interaction process.

It is tempting to compare the values of the ligand translocation barriers with those reported for other charged molecules crossing the core of zwitterionic lipid bilayers. The atomistic value for the forward transition is larger than those typically calculated by atomistic force fields for the translocation of anions. The barrier for $Cl^-$, for example, was estimated as 100.8 $\pm$1.3 kJ/mol in dipalmytoil-phosphatidylcholine (DPPC) by Vorobyov[29] *et al*., and as 98.7 kJ/mol in dimyristoyl-phosphatidylcholine (DMPC) by Khavrutskii[30] *et al*. The CG MPW model, instead, predicts a barrier for $Cl^-$ of 99 kJ/mol (Yesylevskyy [33] *et al*.). The charged side chain of aspartate and glutamate, according to the calculations by MacCallum[45] *et al*., face a barrier of 80 and 85 $\pm$ 1 kJ/mol, whose SM counterpart is about 52 kJ/mol. Overall, the atomistic and CG barriers for ions and charged aminoacids reported in the literature reasonably compare to the result we obtain with the MPW model, while are smaller than those we have calculated at atomistic level.

There are a number of reasons that could account for the discrepancies, at atomistic level, between the translocation barriers of single ions and those of the anionic ligands. First of all, the NP-membrane complex lacks the symmetry with respect to the center of the membrane that is present in the other cases. The anchoring of the charged ligand takes place in a region of the membrane that is deformed by the presence of the NP, and it is hard to make *a priori* considerations about the effect of such deformation on the translocation barrier. Another caveat to the direct comparison of these free energy barriers is the fact that the charged ligand is covalently bound to the NP, which limits its freedom for conformational rearrangements within the membrane core.

The barriers for ligand translocation calculated at CG level by the MPW model are comparable to those of the $Cl^-$ ion. On the one hand this could suggest that the CG model underestimates the effect of the NP on the translocation mechanism. On the other hand, our data indicate that the MPW model reasonably describes the configurational features of the HC state and is able to capture most of the membrane deformations taking place during translocation. Further efforts will be devoted to achieve better statistics at atomistic level and to investigate possible reasons why the translocation of monoatomic ions and charged ligands are characterized by the same energy barriers at CG level.

As a matter of fact, the atomistic barrier of 135 kJ/mol and the CG barrier of 100 kJ/mol we calculated for anchoring a single charged ligand of the NP are extremely large and might prevent the membrane embedding of the NP in realistic conditions. This would be in contrast with experimental evidence of co-localization[11,22] and even complete translocation[11] of anionic Au NPs through dioleolyl-phosphatidylcholine (DOPC) bilayers. We remark that so far we have performed metadynamics calculations using a single reaction coordinate, and thus targeting the same transition mechanisms that occurs spontaneously in unbiased simulations of the NP-membrane interaction with the SM model[18]. We can not rule out the possibility that other transition paths could lead to the final snorkeling configuration at a lower energy cost. Alternative mechanisms that could be possibly explored include the contemporary transition of several ligands, the protonation of the charged ligand terminal during the interaction with the membrane[45], the cooperative effects of more NPs adsorbed at the membrane surface.

**Conclusions**

In this paper we have compared the performance of three alternative versions of the CG Martini model at simulating the interaction between an anionic, monolayer-protected Au NP and a POPC bilayer. The three models are the standard CG Martini model, the same model with the addition of long-range electrostatic interactions, and the polarizable-water Martini model that includes both long-range electrostatics and water polarizability. As a target for assessing the reliability of the three models, we simulated the first stages of the NP-membrane interaction by means of atomistic metadynamics simulations. All the three CG models reasonably reproduce the structural features of the hydrophobic contact configuration, in which the NP stably interacts with one membrane leaflet only. When looking at the transition to the anchored state, in which one charged ligand of the NP translocates through the membrane core to anchor to the distal leaflet, only the MPW model is able to reproduce the translocation mechanism as observed in the atomistic runs. Anchoring happens via formation of ligand-induced water defects in the entrance and in the distal leaflet, sometimes involving the translocation of water molecules across the membrane core. The translocation barriers calculated with the SM and MPME model are highly underestimated, while the MPW model is in reasonable agreement with the atomistic one.

AUTHOR INFORMATION

**Corresponding Author**

*E-mail: rossig@fisica.unige.it.

ACKNOWLEDGMENTS

Giulia Rossi acknowledges funding from the ERC Starting Grant BioMNP – 677513. Part of the calculations was performed at CINECA under the HP10CGNEGB grant.

**Supporting Information**

The atomistic and CG models of the monolayer-protected NP; a comparison between the HC configuration as characterized via unbiased and metadynamics runs; 1D and 2D lipid density profiles in the HC configuration; the description of the disanchoring mechanism as observed in atomistic metadynamics runs. This material is available free of charge via the Internet at http://pubs.acs.org/.

# Supporting Information

# Au nanoparticles in lipid bilayers: a comparison between atomistic and coarse grained models

*Sebastian Salassi[1], Federica Simonelli[1], Davide Bochicchio[1], Riccardo Ferrando[2] and Giulia Rossi\*[1]*

[1]Physics Department, University of Genoa, via Dodecaneso 33, 16146 Genoa, Italy
[2]Chemistry Department, University of Genoa, via Dodecaneso 31, 16146 Genoa, Italy

**Section 1: Atomistic topology**

In this section we report the parameters for the gold core and the ligands for the atomistic model of an anionic nanoparticle (AuNP). The particle is composed of a gold core of 144 Au atoms. 60 ligands are covalently bound to the surface of the AuNP via thiol bonds. Both hydrophobic octanethiol (OT) and hydrophilic mercapto-undecane carboxylate (MUC) can be bound to the core surface. The structure of the functionalized core is the one derived by Lopez-Acevedo *et al.*[1]. The parameterization of the system is developed to be compatible with the OPLS UA force field[2].
The chemical structure of the AuNP is shown in Figure S1.

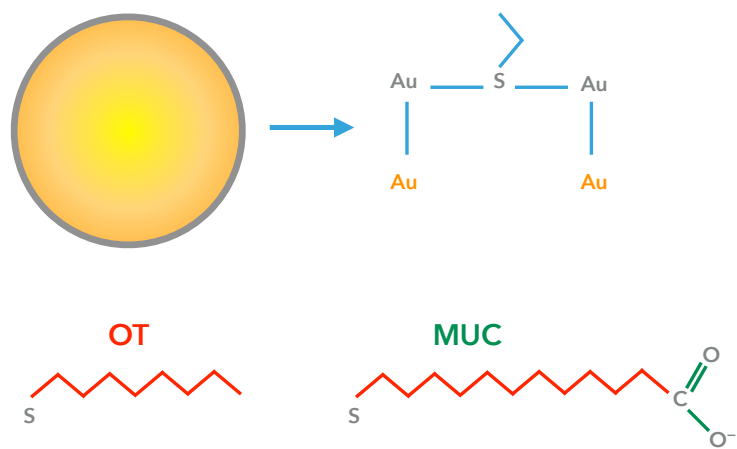

**Figure S1** - Top left: sketch of the core of the AuNP. The contour in grey represents the surface of the core. Top right: schematic representation of the arrangement of the Au and S atoms on the surface of the core in a staple configuration. Bottom left: OT ligand. Red color means hydrophobic character. Bottom right: MUC ligand. Red color means hydrophobic character while green color means hydrophilic character.

*Non-bonded interactions.* The parameterization of the Lennard-Jones parameters $\sigma, \epsilon$ for the Au atom type follows Heinz[3] *et al*.

We assigned the charges of Au atoms as suggested by Jaakko Akola *et al*. who derived them from Bader charge analysis performed on previous calculations by Lopez-Acevedo *et al*.[1].

Au atoms in the core (shown in orange in Figure S1, top) are assigned a slightly positive charge $q_{Au-core} = +0.0286|e|$. Au on the surface (shown in grey in Figure S1, top) are positively charged with $q_{Au-surf} = +0.10273|e|$. The charge of S atoms and of the first $CH_2$ group of each ligand bound to the NP core has been set to neutralize the charge of the gold cluster: S atoms are assigned a negative charge $q_S = -0.12123|e|$. The remaining charge is distributed among the $CH_2$ groups bound to the S atom in each ligand resulting in $q_C = +0.015525|e|$.

A purely repulsive potential is assigned to the interaction between Au and S atoms. If the Lennard-Jones potential is written as

$$U(r) = \frac{C_{12}}{r^{12}} - \frac{C_6}{r^6}$$

then $C_6 = 0$ kJ mol$^{-1}$ nm$^6$ and $C_{12} = 0.92953 \; 10^{-6}$ kJ mol$^{-1}$ nm$^{12}$.

Lennard-Jones parameters and charges for all the atoms in the AuNP are summarized in Table S1.

**Table S1 - Lennard-Jones parameters and charges for the AuNP.**

| Chemical unity | Particle type | $\sigma$ [nm] | $\epsilon$ [kJ mol$^{-1}$] | Charge [$|e|$] |
|---|---|---|---|---|
| Au | Au | 0.2629 | 22.1449 | • Au(core): +0.0286<br>• Au(surface): +0.10273 |
| S | opls_084 | 0.355 | 1.046 | -0.12123 |
| $CH_2$ | opls_071 | 0.3905 | 0.493712 | • 1$^{st}$ of a ligand: +0.015525<br>• others: 0.0 |
| $CH_3$ | opls_068 | 0.3905 | 0.7322 | 0.0 |
| C in COO$^-$ | opls_271 | 0.375 | 0.43932 | 0.7 |
| O in COO$^-$ | opls_272 | 0.296 | 0.87864 | -0.8 |

*Bonded interactions.* We built an elastic network to describe both the gold core and the S shell. Both OPLS for hydrocarbons[4] and AMBER for nucleic acid and proteins[5] force fields were used to derive the bonded parameters for the hydrophobic and hydrophilic parts of the molecules. As for the S–$CH_2$ bond parameters from both Hauptman and Klein[6] and AMBER[5] were used.

Bond, angle and dihedral parameters are summarized in Table S2, S3 and S4, respectively.

The following potentials have been used for the bonded interaction:

1. $$V_b(r) = \frac{1}{2}k^b(r-b)^2 \qquad \text{for bonds}$$

2. $$V_a(\theta) = \frac{1}{2}k^\theta(\theta-\theta^0)^2 \qquad \text{for angles}$$

3. $$V_{rb}(\phi) = \sum_{n=0}^{5} C_n \cos^n(\phi), \qquad \psi = \phi - 180° \qquad \text{for proper dihedrals}$$

4. $$V_d(\phi) = k_\phi(1 + \cos(n\phi - \phi_s)) \qquad \text{for proper/improper dihedrals}$$

Table S2 - Bond parameters for the AuNP.

| Atoms | $b$ [nm] | $k^b$ [kJ mol$^{-1}$ nm$^{-2}$] |
|---|---|---|
| Au–Au (elastic network) | From [1] | 11000 (core); 32500 (surface) |
| S–S (elastic network) | From [1] | 25000 |
| Au–S | From [1] | 32500 |
| S–CH$_2$ | 0.182 [6] | 1.85858 10$^5$ [5] |
| CH$_2$–CH$_2$ | 0.153 [4] | 2.17672 10$^5$ [5] |
| CH$_2$–CH$_3$ | 0.153 [4] | 2.17672 10$^5$ [5] |
| CH$_2$–C | 0.152 [5] | 2.65390 10$^5$ [5] |
| C–O | 0.125 [5] | 5.48941 10$^5$ [5] |

Table S3 - Angle parameters for the AuNP.

| Atoms | $\theta^0$ [deg] | $k^\theta$ [kJ mol$^{-1}$ rad$^{-2}$] |
|---|---|---|
| S–CH$_2$–CH$_2$ | 114.4 [6] | 519.7 [6] |
| | | |
| CH$_2$–CH$_2$–CH$_2$ | 112.0 [4] | 527.4 [5] |
| CH$_2$–CH$_2$–CH$_3$ | 112.0 [4] | 527.4 [5] |
| CH$_2$–CH$_2$–C | 112.0 [5] | 527.4 [5] |
| CH$_2$–C–O | 117.0 [5] | 586.0 [5] |
| O–CH$_2$–O | 126.0 [5] | 669.8 [5] |

Table S4 - Dihedral and improper dihedral parameters for the AuNP.

| Atoms | $C_0$ [kJ mol$^{-1}$] | $C_1$ [kJ mol$^{-1}$] | $C_2$ [kJ mol$^{-1}$] | $C_3$ [kJ mol$^{-1}$] | $C_4$ [kJ mol$^{-1}$] | $C_5$ [kJ mol$^{-1}$] |
|---|---|---|---|---|---|---|
| X–CH$_2$–CH$_2$–X | 8.4015 | 16.7945 | 1.134 | -26.33 | 0.0 | 0.0 |
| | $\phi$ [deg] | $k_\phi$ [kJ mol$^{-1}$] | multiplicity | | | |
| X–CH$_2$–C–X | 180.0 | 0 | 3 | | | |
| Improper | $\phi$ [deg] | $k_\phi$ [kJ mol$^{-1}$] | multiplicity | | | |
| X–O–C–O | 180.0 | 43.953 | 2 | | | |

**Section 2: Coarse-Grained topology**

In this section we summarize the parameters used to model the AuNP at a coarse-grained (CG) level. These parameters were assigned according to the MARTINI force field[7]. The core of the NP is described as in the atomistic case apart from charges which were set to zero for all Au and S atoms in the core. As for the ligands, the OT molecule is made of 2 C1 beads while the MUS (mercapto-undecane sulfonate) ligand is built with 3 C1 and 1 $Q_{da}$ beads with charge $-|e|$. Bonds between MARTINI beads are described with the harmonic potential $V_b(r)$ (see eq. 1. in Section 1) using standard MARTINI parameters $k^b = 1250$ kJ mol$^{-1}$ nm$^{-2}$ and $b = 0.47$ nm. Angle potentials are described by the harmonic function 2 in Section 1 with parameters $k^\theta = 25$ kJ mol$^{-1}$ and $\theta^0 = 180°$. An additional purely repulsive potential has been used for the Au–S interaction with $C_{12} = 0.92953 \cdot 10^{-6}$ kJ mol$^{-1}$ nm$^{12}$. The same repulsive potential was used between Au/S atoms and all other MARTINI beads.

As for the arrangement of the ligands on the NP, we built a central stripe of hydrophobic ligands flanked by two charged poles. A total of 30 hydrophobic and 30 hydrophilic ligands are connected to the core of the NP. This "patched" configuration is shown in Figure S2 (CG model).

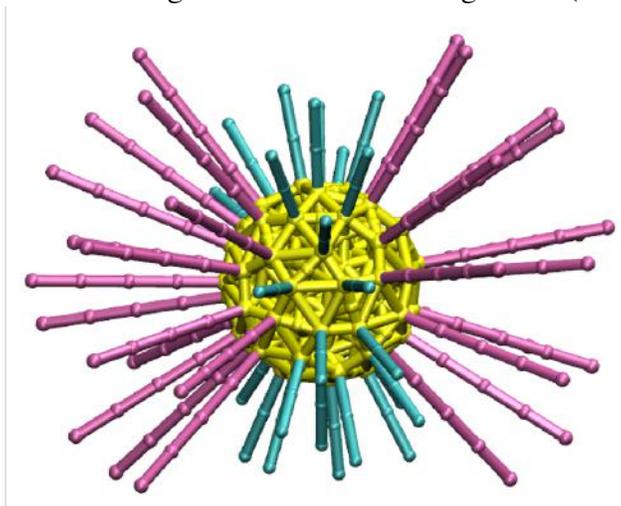

**Figure S2 - Patched NP. In pink the hydrophilic ligands which constitute the two charged poles of the NP, in cyan the hydrophobic ligands of the central stripe. The gold core of the NP is in yellow.**

**Section 3: Initialization of the unbiased MD equilibration runs**

**CG unbiased equilibration runs:** The hydrophobic contact state was obtained from unbiased simulations with the SM force field, which provides spontaneous insertion of the NP from the water phase into the HC configuration. MPME equilibration runs were initialized directly from the configurations obtained with the SM model. For what concerns the equilibration runs performed with the MPW model, the standard Martini water in the box had first to be converted to polarizable water.

**Atomistic unbiased equilibration runs**: As for atomistic simulations, the HC configuration was obtained in this way: (i) the region of an equilibrated membrane with the lowest density of lipid heads (phosphate and choline groups) was selected, (ii) we translated the NP on top of this region, so that the distance between the center of mass of the NP and that of the membrane along the normal to the membrane was about 2 nm. Attention has been paid to the possible overlap of lipids with the NP core, (iii) the system was locally minimized and then equilibrated.

**Section 4: Initialization of the metadynamics runs**

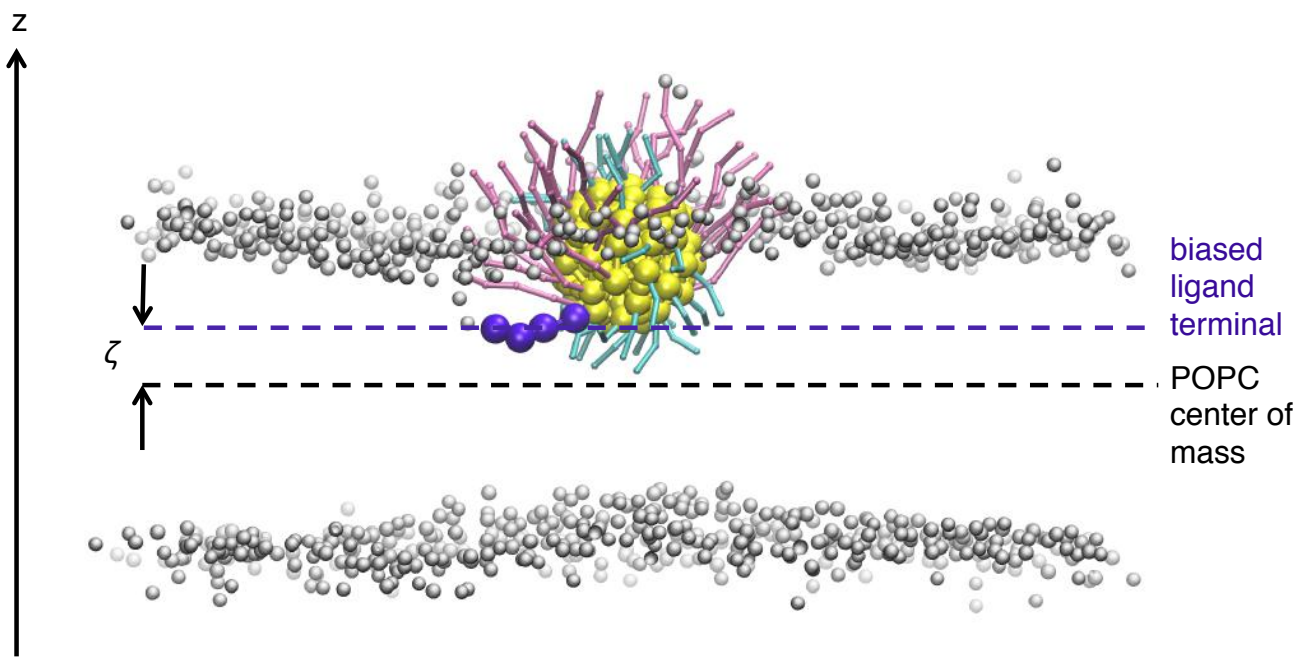

**Figure S3** – After the equilibration of the NP in the HC configuration, the charged ligand to be biased during the metadynamics run was chosen as the one whose covalent link to the NP core had the lowest z coordinate in the initial configuration

**Section 5: The metadynamics bias does not perturb the HC configuration**

In order to understand if biasing the charged ligand terminal with metadynamics can somehow perturb behavior of the system in the HC configuration we have performed some analysis to compare metadynamics runs with unbiased MD simulation. In particular, we studied the total number of contacts

S5

per nanosecond between the charged ligand terminal and water and between the charged ligand terminal and the choline group of the POPC lipids as a function of the z-distance between the COM of the POPC bilayer and the COM of the charged ligand terminal, $\zeta$. The results are shown in Figure S4 for both the atomistic model and the MPW model. Results for unbiased simulations could be obtained only for the region corresponding to the hydrophobic contact state since during unbiased simulations no ligand translocation could be observed.

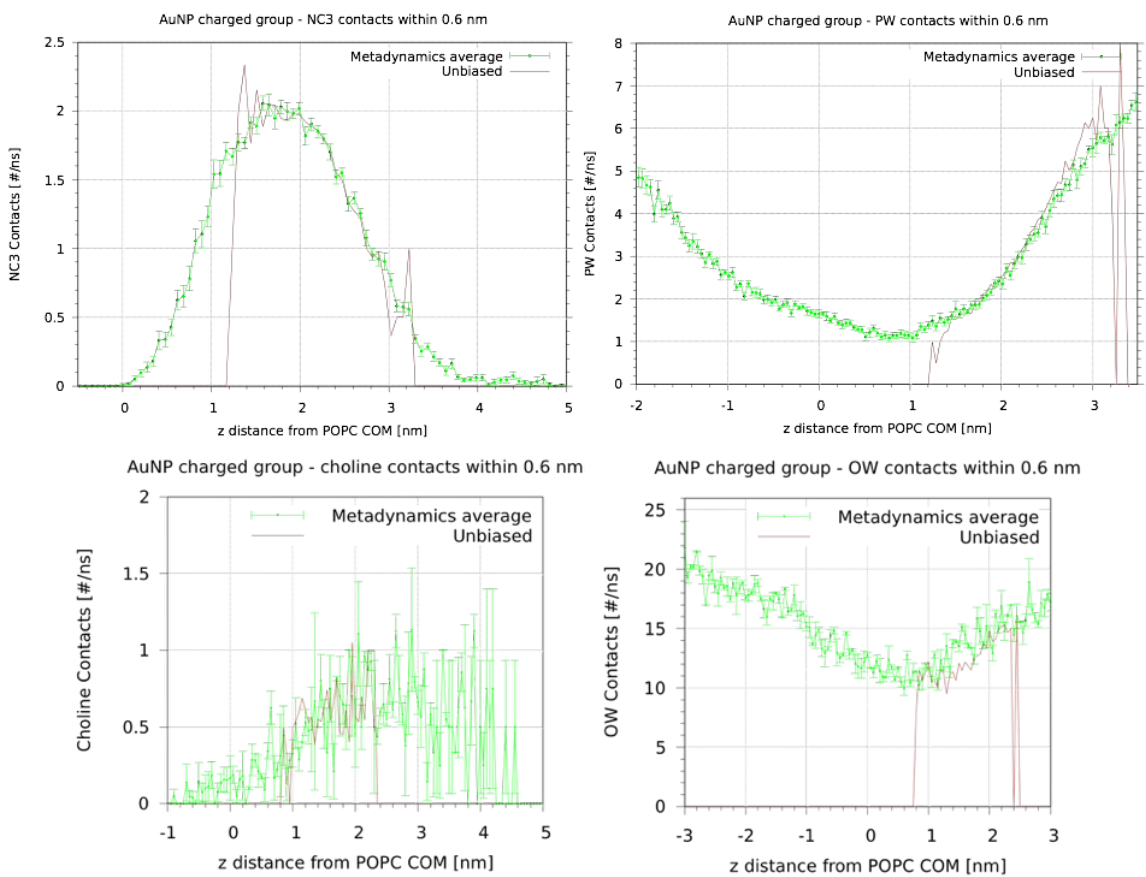

**Figure S4 - Left:** Number of contacts per ns between choline groups of lipids and the charged terminal of one ligand as a function of $\zeta$ for CG PW+PME simulations (top) and atomistic simulations (bottom). **Right:** Number of contacts per ns between water (PW) and the charged terminal of a ligand as a function of $\zeta$ for CG PW+PME (top) and atomistic simulations (bottom). Green curves represent the average of different metadynamics runs (see Table 1 in the main paper) while brown curves refer to one unbiased run.

## Section 6: Identification of forward/backward transitions in metadynamics runs

We considered that the forward transition had been completed as soon as the charged ligand was in contact with the lipid headgroups of the distal leaflet. In other words, we did not use a threshold in the CV space, but rather the visual inspection of (a) the metadynamics trajectory or (b) the time dependence of the CV (see Fig.S5 below). Both provide a very clear indication about the transition from the HC to the anchored state (and viceversa).

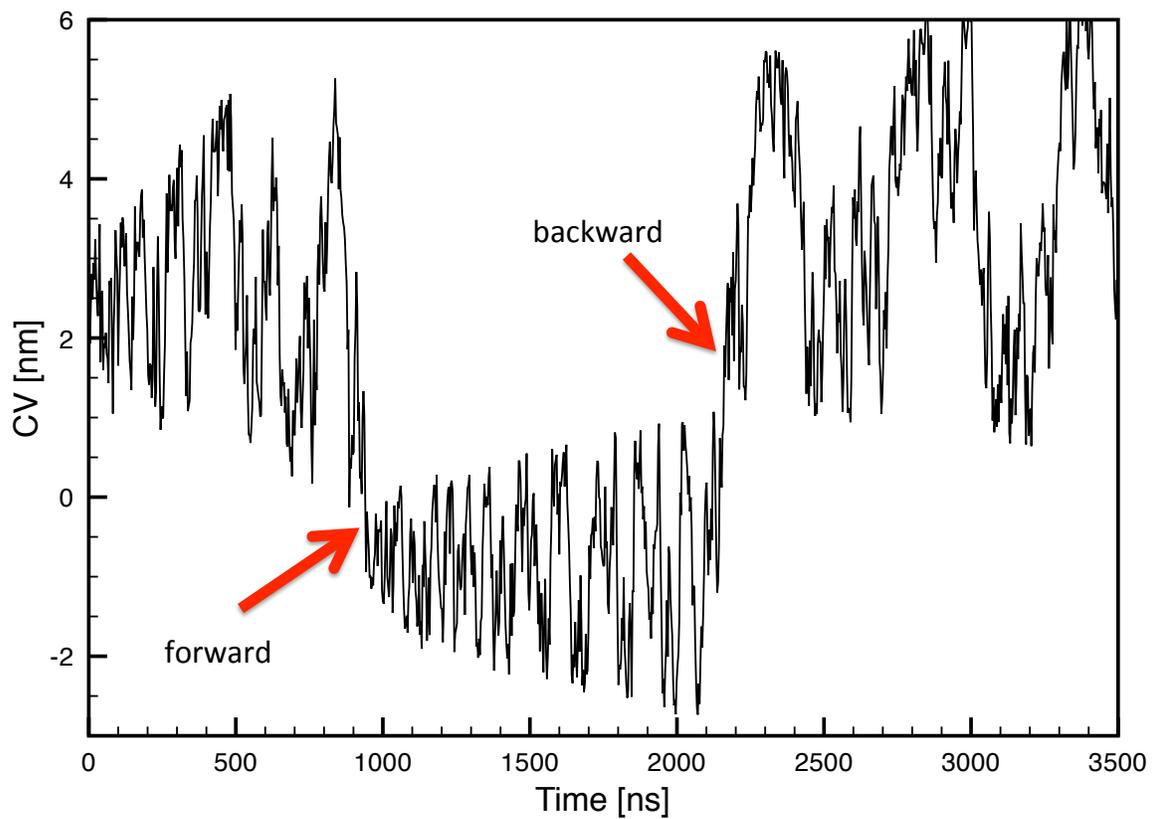

**Fig. S5 – Time dependence of the CV in a metadynamics run completing a full forward/backward cycle. The arrows indicate when the transitions take place, as confirmed by the visual inspection of the trajectories.**

**Section 7: Analysis of the HC configuration from the unbiased MD runs**

**Radial distribution functions (RDF) and maps**

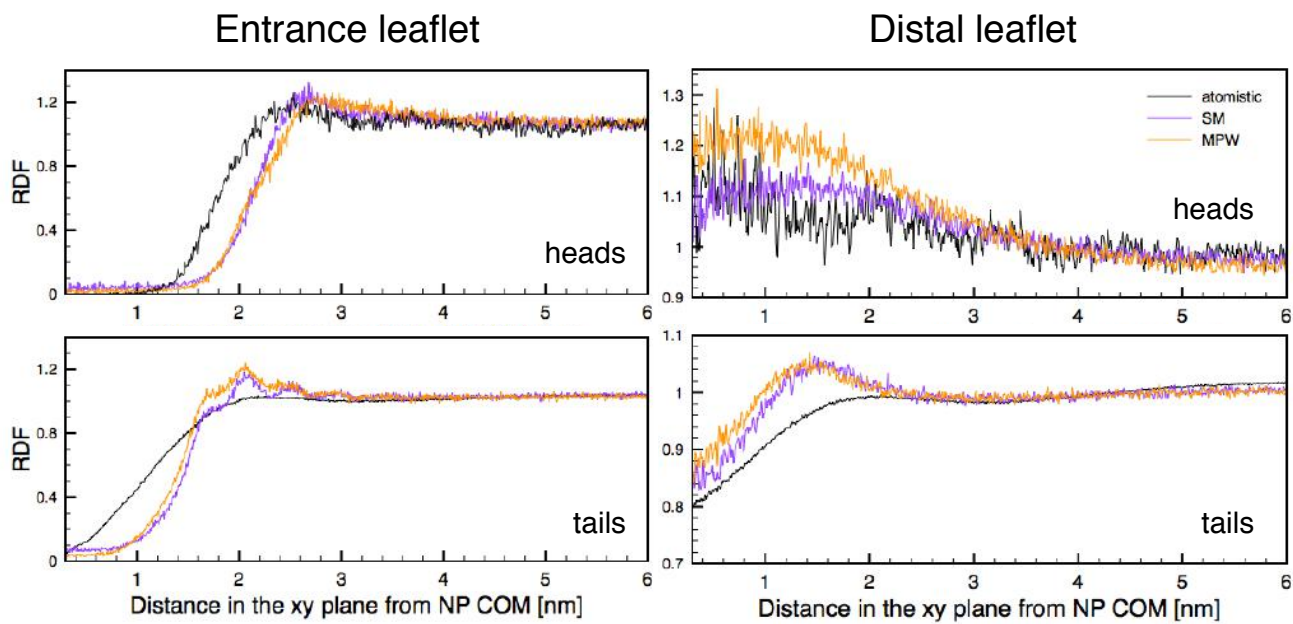

**Figure S6** - RDF of lipid heads and tails as a function of the in-plane distance from the NP center of mass.

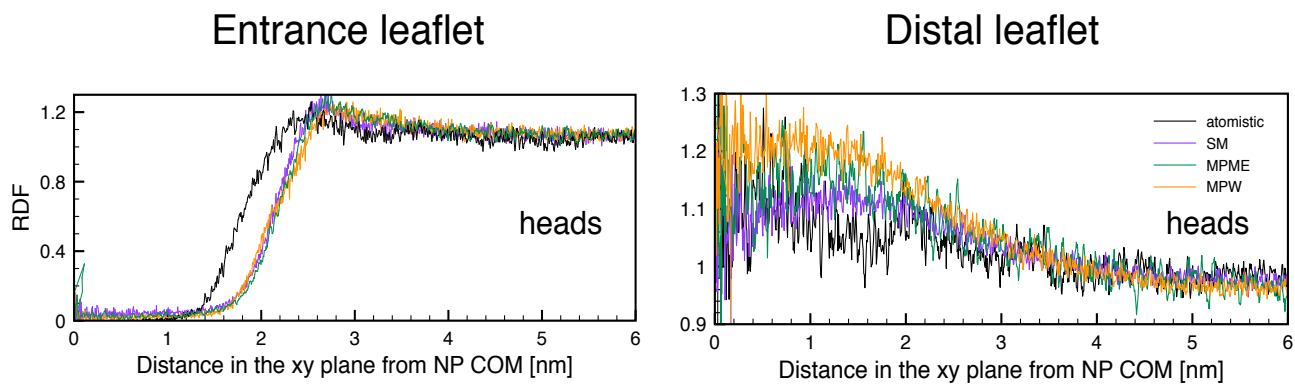

**Figure S6bis** - The same RDFs as in Figure S4 (upper panels), with the addition of the data from the unbiased MD runs with the MPME model (green).

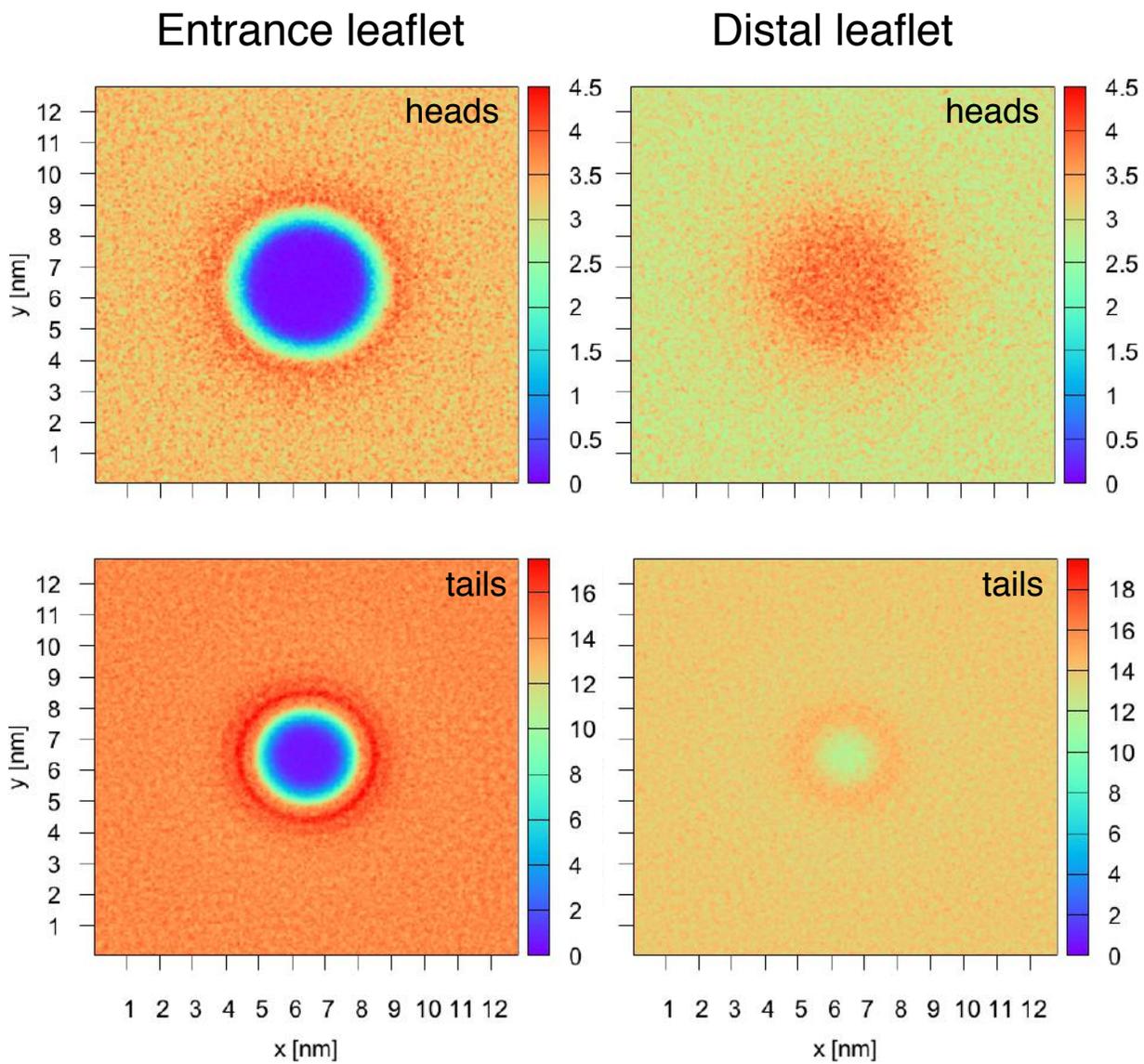

**Figure S7 - Numeric density of lipid heads and tails as obtained with the MPW model.**

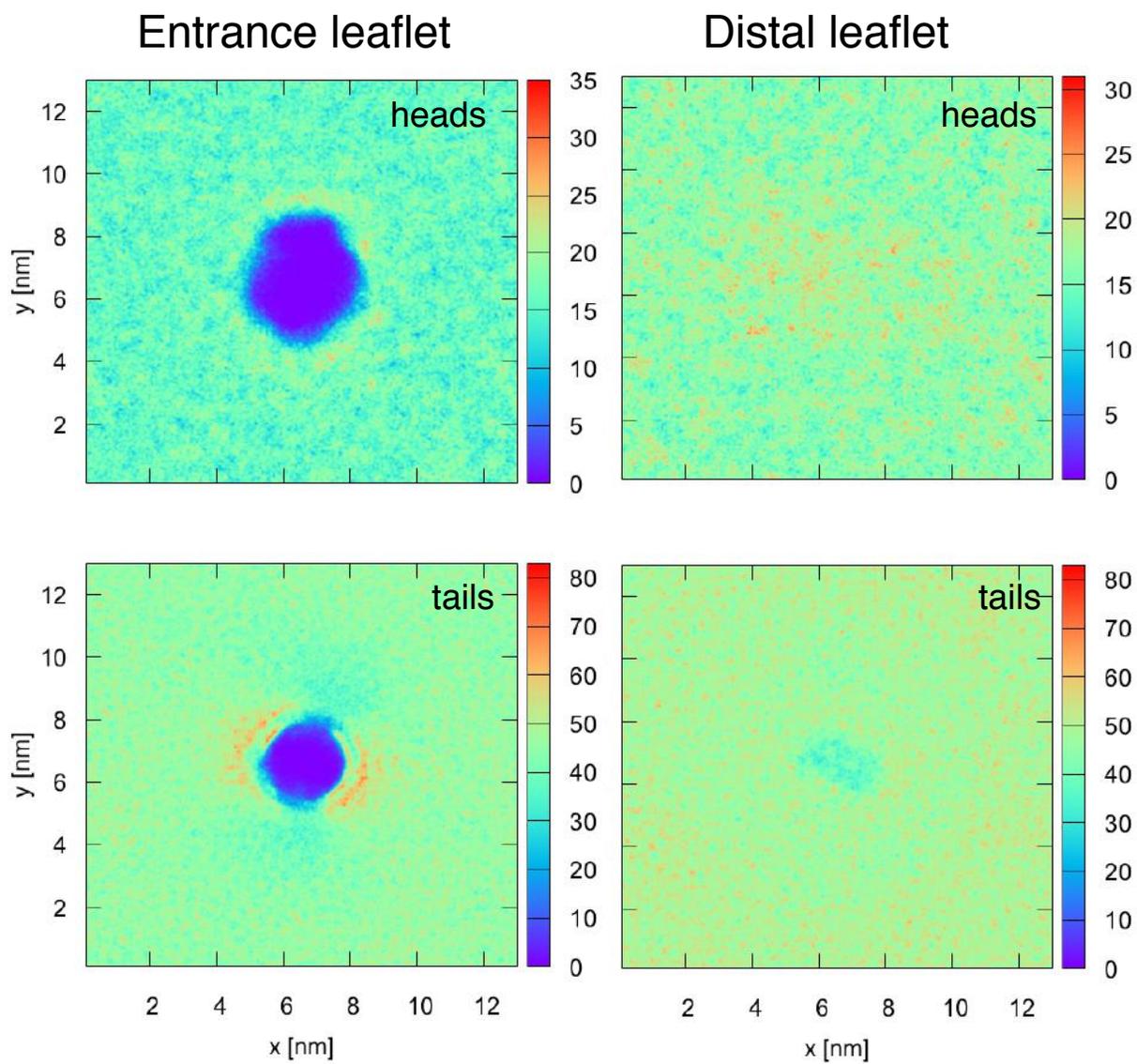

**Figure S8 - Numeric density of lipid heads and tails as obtained with the atomistic model.**

**Section 8: Disanchoring process and water behavior.**

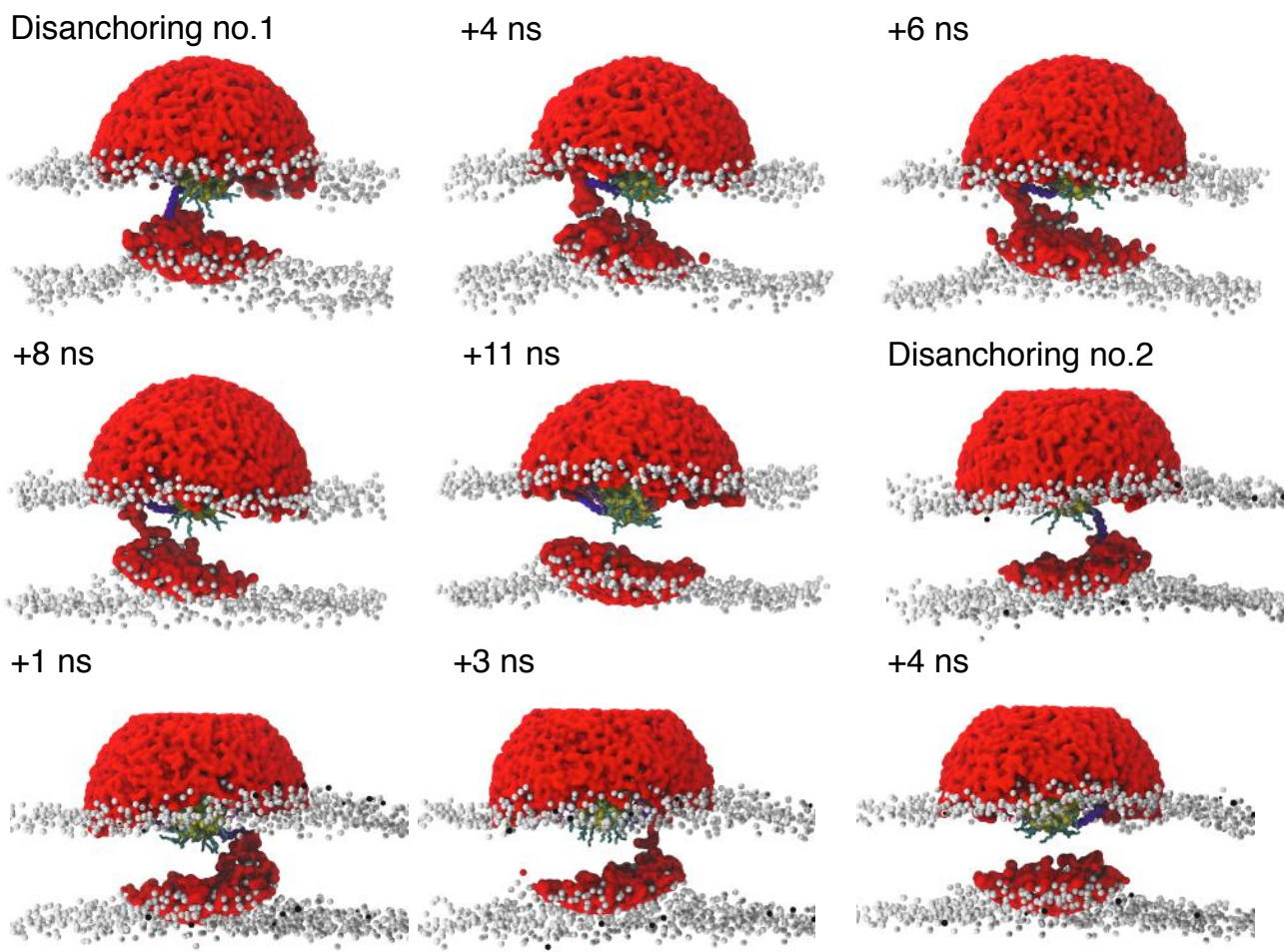

**Figure S9** - Disanchoring mechanisms as observed in our atomistic metadynamics simulations. Lipid heads in grey, tails not shown. Au and S atoms in yellow. Hydrophobic ligands in cyan, anionic ligands in pink. Water molecules within 3.5 nm from the NP center of mass are shown in red. The first disanchoring process takes place via formation of a water pore that is resealed in about 5 ns. In the second disanchoring process, the disanchoring is accompanied by the formation of a single water file whose life time is about 1 ns.